\documentclass[letterpaper,english,american,twocolumn,aps,pre,showpacs,superscriptaddress]{revtex4-1}
\usepackage[LGR,T1]{fontenc}
\usepackage[latin9]{inputenc}
\usepackage{float}
\usepackage{amsmath}
\usepackage{amssymb}
\usepackage{graphicx}

\makeatletter

\pdfpageheight\paperheight
\pdfpagewidth\paperwidth

\ProvideTextCommand{\~}{LGR}[1]{\char126#1}

\usepackage{color}
\usepackage{textcomp}

\usepackage{ulem}
\usepackage{float}
\usepackage{lipsum}
\usepackage{babel}

\makeatother

\usepackage{babel}
\begin{document}

\title{Exact zeros of fidelity in finite-size systems as a signature for
probing quantum phase transitions}
\author{Yumeng Zeng}
\affiliation{Beijing National Laboratory for Condensed Matter Physics, Institute
of Physics, Chinese Academy of Sciences, Beijing 100190, China}
\affiliation{School of Physical Sciences, University of Chinese Academy of Sciences,
Beijing 100049, China }
\author{Bozhen Zhou}
\affiliation{Beijing National Laboratory for Condensed Matter Physics, Institute
of Physics, Chinese Academy of Sciences, Beijing 100190, China}
\author{Shu Chen}
\email{Corresponding author: schen@iphy.ac.cn }

\affiliation{Beijing National Laboratory for Condensed Matter Physics, Institute
of Physics, Chinese Academy of Sciences, Beijing 100190, China}
\affiliation{School of Physical Sciences, University of Chinese Academy of Sciences,
Beijing 100049, China }
\date{\today}
\begin{abstract}
The fidelity is widely used to detect quantum phase transitions, which
is characterized by either a sharp change of fidelity or the divergence
of fidelity susceptibility in the thermodynamical limit when the phase-driving
parameter is across the transition point. In this work, we unveil
that the occurrence of exact zeros of fidelity in
finite-size systems can be applied to detect quantum phase transitions.
In general, the fidelity $\mathcal{F}(\gamma,\tilde{\gamma})$ always
approaches zero in the thermodynamical limit, due to the Anderson
orthogonality catastrophe, no matter whether the parameters of two
ground states ($\gamma$ and $\tilde{\gamma}$) are in the same phase
or different phases, and this makes it difficult to distinguish whether
an exact zero of fidelity exists by finite-size analysis. To overcome
the influence of orthogonality catastrophe, we study finite-size systems
with twist boundary conditions, which can be introduced by applying
a magnetic flux, and demonstrate that exact zeros
of fidelity can be always accessed by tuning the magnetic flux when
$\gamma$ and $\tilde{\gamma}$ belong to different phases. On the
other hand, no exact zero of fidelity can be observed if $\gamma$
and $\tilde{\gamma}$ are in the same phase. We demonstrate the applicability
of our theoretical scheme by studying concrete examples, including
the Su-Schrieffer-Heeger model, Creutz model and Haldane model. Our
work provides a practicable way to detect quantum phase transitions
via the calculation of fidelity of finite-size systems.
\end{abstract}
\maketitle

\section{Introduction}

A quantum phase transition represents qualitative changes in the ground
state properties of quantum systems induced by varying control parameters
\cite{Sachdev}. These changes become nonanalytic around the phase
transition point in the infinite-size limit and can be revealed by
the response of physical quantities to changes in phase-driving parameters.
The notion of fidelity between two pure ground states has become a
well-established method of detection of quantum phase transitions
\cite{PZ2006PRE,Zanardi,Zanardi2,GuSJ}. So far, fidelity and fidelity
susceptibility have been applied to study phase transitions in various
systems \cite{PZ2006PRE,Zanardi,Zanardi2,GuSJ,SC2008PRA,QuanHT,Damski,Vezzani,SunG2015,SunG,YangMF,ZhouHQ,SC2007,SJGu2008,WangXQ,Konig,Dutta}.
Given $\gamma$ is a phase-driving parameter of the system, the fidelity
is defined as the module of the overlap of two ground states, i.e.,
\begin{equation}
\mathcal{F}(\gamma,\tilde{\gamma})=\left|\left\langle \psi^{0}(\gamma)\right|\left.\psi^{0}(\tilde{\gamma})\right\rangle \right|,
\end{equation}
where $\left|\psi^{0}(\gamma)\right\rangle $ is the ground state
of the Hamiltonian $H(\gamma)$ and $\tilde{\gamma}$ is a parameter
value different from $\gamma$. If two parameters are very close,
i.e., $\delta=\tilde{\gamma}-\gamma$ is a small quantity, quantum
phase transitions are associated with a drop in the fidelity when
$\gamma$ and $\tilde{\gamma}$ are in different phases. Hence the
second derivative of the fidelity, called as fidelity susceptibility,
is divergent at the quantum phase transition point in the thermodynamical
limit.

The divergence of fidelity susceptibility at quantum phase transition
point suggests that $\mathcal{F}(\gamma,\tilde{\gamma})$ may have
a different property when $\gamma$ and $\tilde{\gamma}$ are in different
phase regions. Recent studies on dynamical quantum phase transitions
\cite{Heyl2013PRL,Budich2016PRB,Heyl2018RPP,Dora,YangC,Heyl2015PRL,Liska,ZengYM,ZhouBZ2021,Karrasch2013PRB}
have unveiled that the Loschmodt echo can have exact zero points at
some critical times when the post-quench parameter and pre-quench
parameter are in different phase regions \cite{Heyl2013PRL,Budich2016PRB,Dora,Liska,ZengYM,ZhouBZ2021,WuLA},
whereas it may have no zero point if the post-quench parameter and
pre-quench parameter are in the same phase region. Motivated by these
progresses, one may ask whether $\mathcal{F}(\gamma,\tilde{\gamma})$
has (has no) exact zero points when $\gamma$ and $\tilde{\gamma}$
are in different (the same) phase regions? Although the question seems
very natural, it is hard to give a simple answer due to the existence
of the Anderson orthogonality catastrophe (OC) in the thermodynamical
limit. According to the OC, when the system size increases to infinity,
even a slight perturbation leads to a many-body ground state having
zero overlap with the slightly perturbed state \cite{OC}. Therefore,
$\mathcal{F}$ approaches zero in the thermodynamical limit as long
as $\tilde{\gamma}$ and $\gamma$ are not equal, no matter whether
$\tilde{\gamma}$ and $\gamma$ are in the same or different phase
regions.

In this work, we unveil that the fidelity exhibits different features
for $\tilde{\gamma}$ and $\gamma$ in the same and different phase
regions by studying several typical two-band quantum systems, which
display quantum phase transitions. For systems with translation symmetry,
the fidelity can be represented as the product of $\mathcal{F}_{k}$, i.e.,
$\mathcal{F}(\gamma,\tilde{\gamma})=\prod_{k}\mathcal{F}_{k}(\gamma,\tilde{\gamma})$, where $\mathcal{F}_{k}(\gamma,\tilde{\gamma})$
is the fidelity of the $k$-mode, which represents the overlap of wavefuntions in the momentum space with momentum $k$ and different parameters $\gamma$ and $\tilde{\gamma}$.
Although the fidelity is found
to decay exponentially with the increase of $L$ as
a result of the OC and thus always approaches zero in the thermodynamical
limit, we demonstrate that there exists at least one
$k_{c}$-mode so that
$\mathcal{F}_{k_{c}}=0$ is fulfilled for $\tilde{\gamma}$ and $\gamma$ in different
phases, whereas such a $k_{c}$-mode is absent for $\tilde{\gamma}$
and $\gamma$ in the same phase.
Under the periodic boundary condition (PBC), the momentum $k$ of a finite-size system can take only some discrete values, which do not cover the whole momentum space continuously. Therefore, usually the $k_{c}$-mode
is only accessible in the thermodynamical limit,  and thus the exact zero of the fidelity
does not exist in finite-size systems. For the modes $k'$ located in the vicinity of $k_{c}$,  $\mathcal{F}_{k'}$ approach zero in terms of $\mathcal{F}_{k'}\propto1/L$
when $\tilde{\gamma}$ and $\gamma$ are in different
phases. Due to the existence of OC, it is hard to distinguish whether the $k_c$ modes exist from the analysis of size-dependent behavior of $\mathcal{F}(\gamma,\tilde{\gamma})$, as $\mathcal{F}(\gamma,\tilde{\gamma})$ always decays exponentially with the increase of $L$ no matter whether $\gamma$ and $\tilde{\gamma}$ in the same phase or different phases. In order to reduce the impact of OC, we shall fix the lattice size $L$ and  introduce a magnetic
flux $\phi$ into the periodic boundary system, so we can shift the momentum $k$ continuously to access $k_c$ by choosing a proper
twist boundary condition. In this way, we can always
access $\mathcal{F}(\gamma,\tilde{\gamma})=0$ by tuning $\phi$, when $\tilde{\gamma}$
and $\gamma$ are in different phase regions, whereas no exact zero of fidelity is accessible for $\tilde{\gamma}$
and $\gamma$ in the same phase region. Therefore, under a proper
twist boundary condition, the fidelity $\mathcal{F}(\gamma,\tilde{\gamma})$
of a finite-size system does not change continuously with $\tilde{\gamma}$
and a discontinuous change from nonzero to zero value occurs at the
phase transition point. As a consequence, the discontinuity of the
fidelity can be viewed as a signature for detecting quantum phase
transition in finite-size systems.

In order to reduce the influence of the system size, it is convenient
to introduce a decay rate function defined by
\begin{equation}
\alpha=-\frac{1}{L}\ln\mathcal{F}(\gamma,\tilde{\gamma}).
\end{equation}
For a finite $L$, the exact zero of the fidelity means the divergence
of $\alpha$. Therefore, we can observe the divergence of $\alpha$
by tuning $\phi$ when $\tilde{\gamma}$ and $\gamma$ are in different
phase regions, whereas no divergence of $\alpha$ can be observed
for the case with $\tilde{\gamma}$ and $\gamma$ in the same phase
region. This indicates that the emergence of singularity in the decay
rate function of a finite-size system via the modulation of $\phi$
can be used to detect quantum phase transition.

\section{Models, scheme and results}

Consider a general two-band system with the Hamiltonian in momentum
space described by
\begin{equation}
\hat{h}_{k}(\gamma)=\sum_{{\beta}=x,y,z}d_{{\beta},k}(\gamma)\hat{\sigma}_{{\beta}}+d_{0,k}(\gamma)\hat{\mathbb{I}},\label{eq:hk}
\end{equation}
where $\hat{h}_{k}(\gamma)$ is the Hamiltonian of $k$-mode with
momentum $k$; $\gamma$ is a phase transition driving parameter;
$\hat{\sigma}_{{\beta}}$ (${\beta}=x,y,z$)
are Pauli matrices; $d_{{\beta},k}(\gamma)$ and $d_{0,k}(\gamma)$
are the corresponding vector components of $\hat{h}_{k}(\gamma)$;
and $\hat{\mathbb{I}}$ denotes the unit matrix. The fidelity of the
system can be represented as
\begin{equation}
\mathcal{F}(\gamma,\tilde{\gamma})=\prod_{k}\mathcal{F}_{k}=\prod_{k}\left|\left\langle \psi_{k}^{0}(\gamma)\right|\left.\psi_{k}^{0}(\tilde{\gamma})\right\rangle \right|,
\end{equation}
where $|\psi_{k}^{0}(\gamma)\rangle$ and $|\psi_{k}^{0}(\tilde{\gamma})\rangle$
are the ground state of $\hat{h}_{k}(\gamma)$ and $\hat{h}_{k}(\tilde{\gamma})$,
respectively. Then we have
\begin{equation}
\mathcal{F}_{k}=\sqrt{\frac{\underset{{\beta}}{\sum}d_{{\beta},k}(\gamma)d_{{\beta},k}(\tilde{\gamma})}{2E_{k}\tilde{E}_{k}}+\frac{1}{2}},
\end{equation}
where $E_{k}=\sqrt{\underset{{\beta}}{\sum}d_{{\beta},k}^{2}(\gamma)}$
and $\tilde{E}_{k}=\sqrt{\underset{{\beta}}{\sum}d_{{\beta},k}^{2}(\tilde{\gamma})}$.

To ensure $\mathcal{F}=0$, one needs at least one $k$-mode fulfilling
$\mathcal{F}_{k}=0$, which gives rise to the following four constraint
relations
\begin{eqnarray}
d_{x,k}(\gamma)d_{y,k}(\tilde{\gamma})=d_{y,k}(\gamma)d_{x,k}(\tilde{\gamma}),\label{eq:constraint1}\\
d_{x,k}(\gamma)d_{z,k}(\tilde{\gamma})=d_{z,k}(\gamma)d_{x,k}(\tilde{\gamma}),\label{eq:constraint2}\\
d_{y,k}(\gamma)d_{z,k}(\tilde{\gamma})=d_{z,k}(\gamma)d_{y,k}(\tilde{\gamma}),\label{eq:constraint3}\\
\underset{{\beta}}{\sum}d_{{\beta},k}(\gamma)d_{{\beta},k}(\tilde{\gamma})<0.\label{eq:constraint4}
\end{eqnarray}
The first three equations determine the value of $k_{c}$, and the
last one determines the phase transition point $\gamma_{c}$. The
four formulas should be satisfied simultaneously. It means that $\vec{d}_{k}(\gamma)$
and $\vec{d}_{k}(\tilde{\gamma})$ should be antiparallel on the bloch
sphere. Note that here $\tilde{\gamma}$ and $\gamma$ should be in
two adjacent phases.

To make our discussion concrete, firstly we consider the Su-Schrieffer-Heeger
(SSH) model as a showcase example and give the details of calculation.
Then we generalize our study to the Creutz model and Haldane model.

\subsection{SSH model}

The SSH model \cite{SSH} is described by the Hamiltonian
\begin{equation}
 H=\sum_{j=1}^{L}(t_{1}c_{j,A}^{\dagger}c_{j,B}+t_{2}c_{j,B}^{\dagger}c_{j+1,A}+\mathrm{H.c.}),
\end{equation}
where $t_{1}$ and $t_{2}$ denote the intracellular and intercellular
hopping amplitudes respectively, and $c_{j,A(B)}^{\dagger}$
and $c_{j,A(B)}$ are fermionic creation and annihilation operators
of the $A(B)$ sublattice on the $j$-th site. By taking
a Fourier transformation $c {\color{red}^\dagger}_{j,A(B)}= \frac{1}{\sqrt{L}} \sum_{k} e^{ikj} c {\color{red}^\dagger}_{k,A(B)}$, the Hamiltonian in the momentum space can be written as
\begin{equation}
H= \sum_{k} \psi_k^{\dagger} \hat{h}_k \psi_k,
\end{equation}
where $\psi_k =(c_{k,A},c_{k,B})^{T}$ and $\hat{h}_k$ takes the form of Eq.(\ref{eq:hk})
with the vector components
\begin{eqnarray}
d_{x,k} & = & t_{1}+t_{2}\cos k,\\
d_{y,k} & = & -t_{2}\sin k,
\end{eqnarray}
and $d_{z,k}=d_{0,k}=0$. The SSH model possesses two topologically
different phases for $t_{2}>t_{1}$ and $t_{2}<t_{1}$ with a phase
transition point at $t_{2c}/t_{1}=1$ \cite{LiLH2014}. Setting $\gamma=t_{2}/t_{1}$,
the fidelity of the SSH model is
\begin{equation}
\mathcal{F}_{k}=\sqrt{\frac{1+(\gamma+\tilde{\gamma})\cos k+\gamma\tilde{\gamma}}{2\sqrt{(1+2\gamma\cos k+\gamma^{2})(1+2\tilde{\gamma}\cos k+\tilde{\gamma}^{2})}}+\frac{1}{2}}.
\end{equation}
According to Eqs. (\ref{eq:constraint1})-(\ref{eq:constraint4}),
the constraint relations for the occurrence of exact zeros of $\mathcal{F}$
are
\begin{equation}
\begin{cases}
(\gamma+1)(\tilde{\gamma}+1)<0, & k_{c}=0;\\
(\gamma-1)(\tilde{\gamma}-1)<0, & k_{c}=\pi.
\end{cases}\label{eq:gamma}
\end{equation}

For a finite-size system under the PBC, the momentum $k$ takes discrete
values $k=2\pi m/L$ with $m=-L/2+1,-L/2+2,\cdots,L/2$ if $L$ is
even or $m=-(L-1)/2,-(L-1)/2+1,\cdots,(L-1)/2$ if $L$ is odd. For
the case of $\gamma>0$ and $\tilde{\gamma}>0$, if $\gamma<1$, Eq.
(\ref{eq:gamma}) is fulfilled only for $\tilde{\gamma}>1$; if $\gamma>1$,
Eq. (\ref{eq:gamma}) is fulfilled only for $\tilde{\gamma}<1$. It
means that $\mathcal{F}=0$ happens only when $\gamma$ and $\tilde{\gamma}$
are in different topological phases which are separated by the phase
transition point $\gamma_{c}=1$ and the corresponding $k$-mode is
$k_{c}=\pi$.
\begin{figure}
\begin{centering}
\includegraphics[scale=0.45]{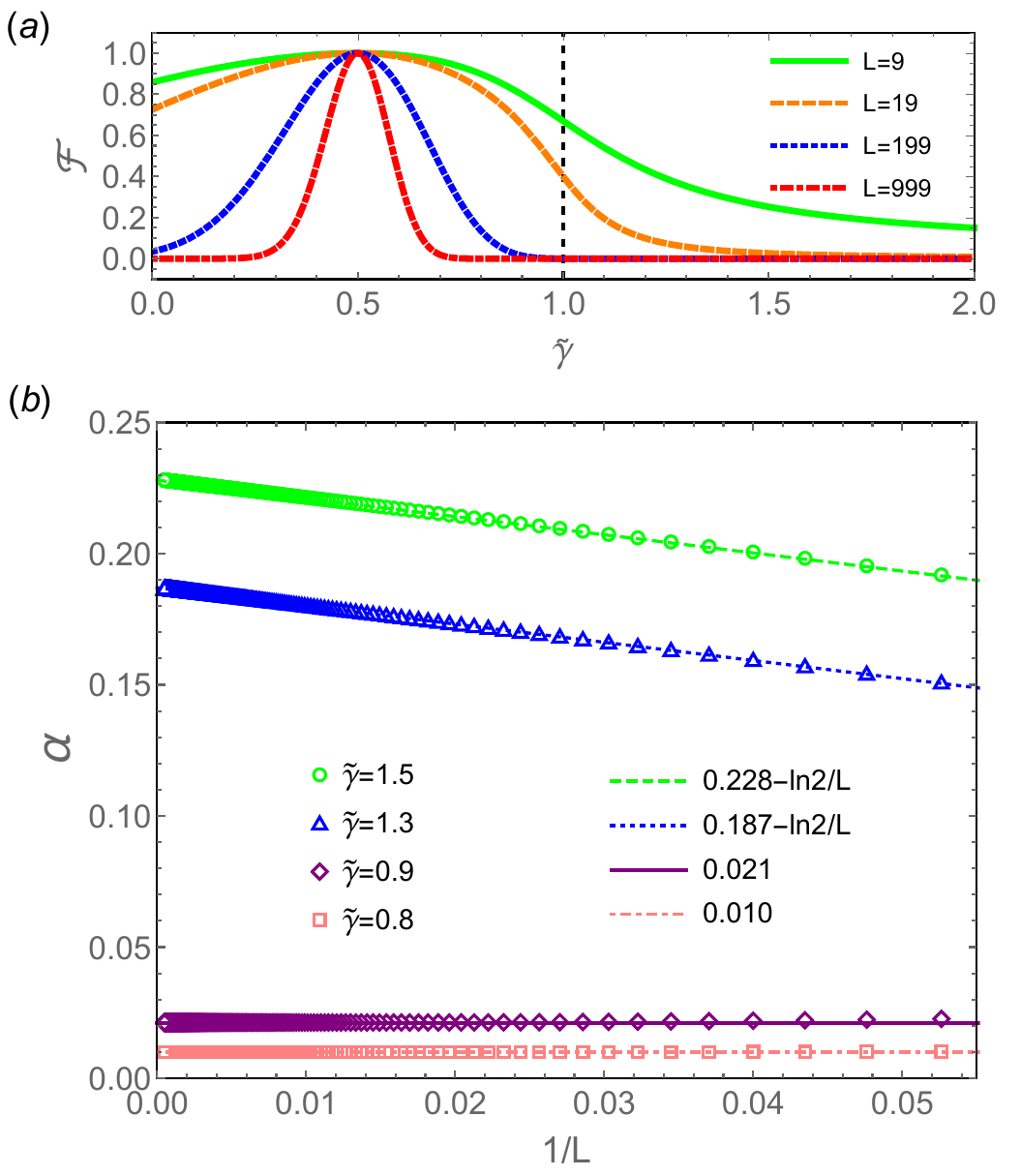}
\par\end{centering}
\caption{(a) The fidelity $\mathcal{F}$ of the SSH model versus $\tilde{\gamma}$
for different system sizes $L=9$, $19$, $199,$ and $999$. The
vertical dashed line guides the critical point $\tilde{\gamma}_{c}=1$.
(b) The image of $\alpha$ versus $1/L$. The pink squares, purple diamonds, blue triangles and
green circles denote results for $\tilde{\gamma}=0.8,\ 0.9,\ 1.3$ and
$1.5$, respectively. The dot-dashed pink, solid purple, dotted blue and
dashed green lines represent the corresponding
fitting lines. Here we take $\gamma=0.5$ and use the periodic boundary condition. \label{Fig1}}
\end{figure}

Now we consider the case with odd size $L$ under PBC. The momentum
$k$ enforced by the PBC can take the value of $k=\pi(1-1/L)$, which
approaches $k_{c}=\pi$ in the limit of $L\rightarrow\infty$. In
Fig. \ref{Fig1}(a), we display $\mathcal{F}$ versus $\tilde{\gamma}$
by fixing $\gamma=0.5$ for various $L$. It shows that $\mathcal{F}$
is finite for small odd sizes and approaches zero with the increase
of $L$ for $\tilde{\gamma}$ and $\gamma$ either in the same phase
or different phases. In Fig. \ref{Fig1}(b), we plot the image of
$\alpha$ versus $1/L$. While $\alpha$ is constant for $\tilde{\gamma}$
and $\gamma$ in the same phase, $\alpha$ versus $1/L$ can be well
fitted by an oblique line with the slope $-\ln2$ for $\tilde{\gamma}$
and $\gamma$ in different phases, i.e., the fitting curves for $\tilde{\gamma}$
and $\gamma$ in the same  or different phase are described by $-\frac{1}{L}\ln \mathcal{F} = c$ or $-\frac{1}{L}\ln \mathcal{F} = c-\frac{1}{L} \ln2$, respectively, where $c$ represents a constant. The additional term of $-\frac{1}{L} \ln2$ is originated from the
existence of the $k_c$-mode ($\mathcal{F}_{k_{c}}=0$) \cite{note}.


\begin{figure}
\begin{centering}
\includegraphics[scale=0.5]{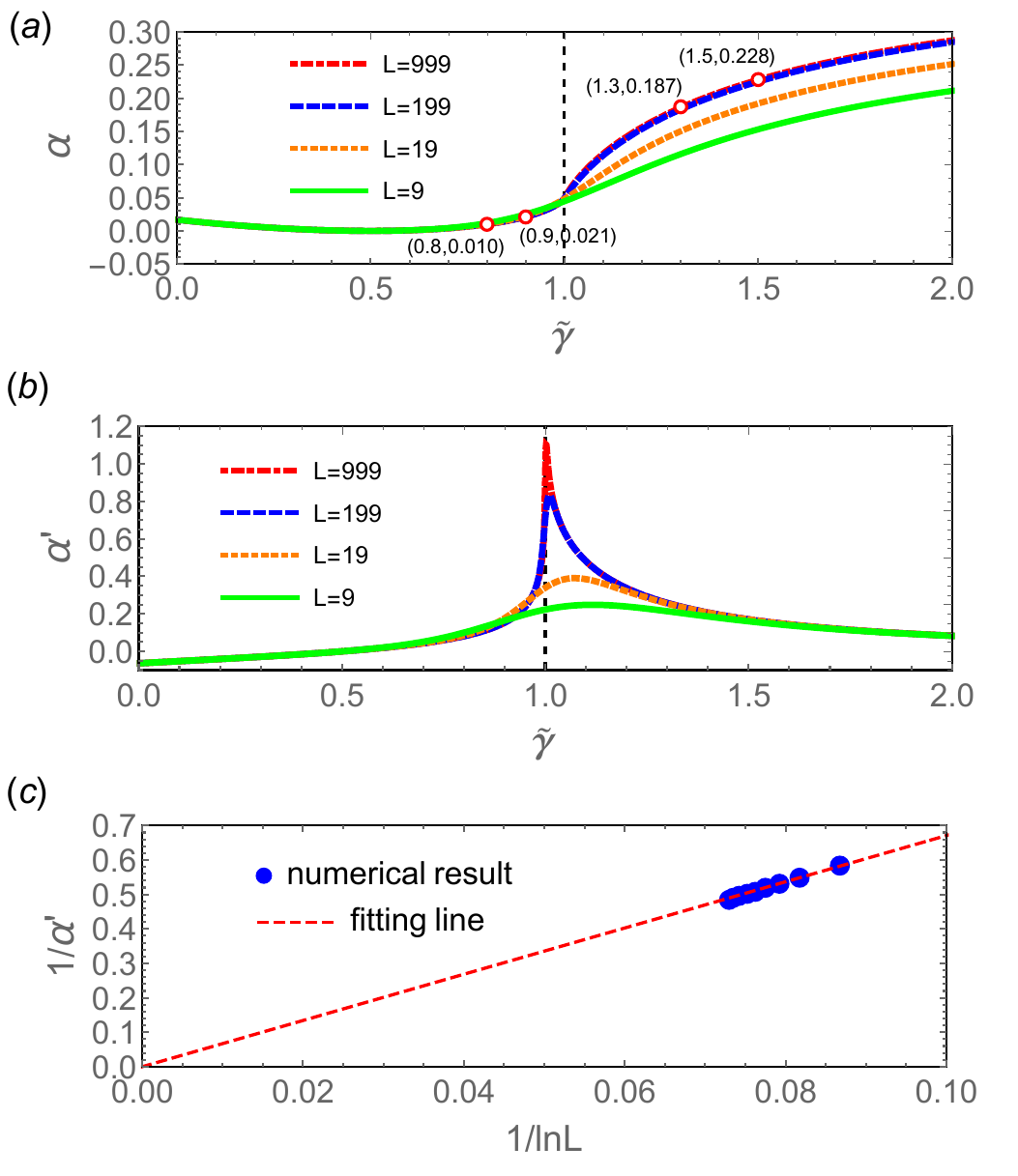}
\par\end{centering}
\caption{(a) The images of $\alpha$ versus $\tilde{\gamma}$ for the SSH model
with different system sizes $L=9$, $19$, $199$, and $999$. The
four red circles denote values of $\alpha$ in the thermodynamical
limit obtained from finite-size analysis for $\tilde{\gamma}=0.8,\ 0.9,\ 1.3$,
and $1.5$, respectively. (b) The images of the derivative of $\alpha$
versus $\tilde{\gamma}$ for the SSH model with different system sizes
$L=9$, $19$, $199$, and $999$. The vertical dashed line guides
the critical point $\tilde{\gamma}_{c}=1$. (c) The
image of the inverse of the derivative of $\alpha$ versus $1/\ln L$
for $\tilde{\gamma}=1$. The blue points are numerical result of $L\in[100000,1000000]$.
The red dashed line is a fitting line. Here we take $\gamma=0.5$ and use the periodic boundary condition.
\label{Fig2}}
\end{figure}

In Fig. \ref{Fig2}(a), we plot $\alpha$ versus $\tilde{\gamma}$
for the SSH model with fixed $\gamma=0.5$ and different system sizes
$L=9$, $19$, $199,$ and $999$. For $\tilde{\gamma}$ and $\gamma$
in the same phase, $\alpha$ is almost the same for different sizes.
For $\tilde{\gamma}$ and $\gamma$ in different phases, the value
of $\alpha$ goes up and tends to a fixed value with the increase
of $L$,\textcolor{red}{{} }which is in accordance with the result of
Fig. \ref{Fig1}(b). Given $\gamma=0.5$, for the case of $\tilde{\gamma}<1$,
$\mathcal{F}\rightarrow0$ in the limit of $L\rightarrow\infty$ is
purely caused by the OC. However, for the case of $\tilde{\gamma}>1$,
besides the OC, $k_{c}$ can be approached in terms of $\min|k-k_{c}|=\pi/L$,
thus $\mathcal{F}_{k_{c}}=0$ is achievable in the thermodynamical
limit of $L\rightarrow\infty$. In contrast, no $k$-mode fulfills
$\mathcal{F}_{k}=0$ for the case of $\tilde{\gamma}<1$. The different
scaling behaviors of the decay rate function $\alpha$ in the regions
of $\tilde{\gamma}<1$ and $\tilde{\gamma}>1$ are due to the absence
and presence of the $k_{c}$ mode in these regions.

To explore the non-analytical behavior of $\alpha(\tilde{\gamma})$
at the critical point, we calculate the derivative of $\alpha(\tilde{\gamma})$
with respect to $\tilde{\gamma}$ via
\begin{equation}
\alpha'(\tilde{\gamma})=-\frac{1}{2L}\sum_{k}\left[\frac{E_{k}\tilde{E}'_{k}+\underset{\beta}{\sum}d_{\beta,k}(\gamma)d'_{\beta,k}(\tilde{\gamma})}{E_{k}\tilde{E}_{k}+\underset{\beta}{\sum}d_{\beta,k}(\gamma)d_{\beta,k}(\tilde{\gamma})}-\frac{\tilde{E}'_{k}}{\tilde{E}_{k}}\right].
\end{equation}
Here $\tilde{E}'_{k}$ and $d'_{\beta,k}(\tilde{\gamma})$
are derivatives of $\tilde{E}_{k}$ and $d_{\beta,k}(\tilde{\gamma})$
with respect to $\tilde{\gamma}$, respectively.
In Fig. \ref{Fig2}(b), we plot $\alpha'$ versus $\tilde{\gamma}$
for the SSH model with different system sizes $L=9$, $19$, $199,$
and $999$. Fig. \ref{Fig2}(c) is the result of finite-size-scaling
analysis of $\alpha'$ for $\gamma=0.5$ and $\tilde{\gamma}=1$.
It shows that $\alpha'$ gradually diverges at the critical point
$\gamma_{c}=1$ as $L$ increases. Therefore, $\alpha$ is nonanalytic
at the critical point $\gamma_{c}=1$ in the thermodynamical limit.

Next we unveil that $\mathcal{F}=0$ can be realized even in a finite-size
system for $\tilde{\gamma}$ and $\gamma$ in different phases if
we introduce a magnetic flux $\phi$ into the system. The effect of
the magnetic flux is equivalent to the introduction of a twist boundary
condition in real space $c_{L+1,A(B)}^{\dagger}=c_{1,A(B)}^{\dagger}e^{i\phi}(\phi\in(0,2\pi))$.
Under the twist boundary condition, the quantized momentum $k=\frac{2\pi m+\phi}{L}$
is shifted by a factor $\phi/L$, where $m=-L/2+1,-L/2+2,\ldots,L/2$
for an even $L$ or $m=-(L-1)/2,-(L-1)/2+1,\ldots,(L-1)/2$ for an
odd $L$. Therefore, for an odd $L$ we can always achieve $k_{c}$
by tuning the flux $\phi$ to $\phi_{c}=\pi$. Let $\Delta=\left|\phi_{c}-\phi\right|$,
we can get
\begin{equation}
\alpha=-\frac{1}{L}(\ln\mathcal{F}_{k^{*}}+\sum_{k\neq k^{*}}\ln\mathcal{F}_{k}),
\end{equation}
where $\mathcal{F}_{k^{*}}$ is the $k^{*}$-mode which is closest
to $k_{c}$, i.e. $\left|k^{*}-k_{c}\right|=\Delta/L$. Let $\Delta\rightarrow0$,
we can get
\begin{equation}
\mathcal{F}_{k^{*}}\approx\frac{\left|\tilde{\gamma}-\gamma\right|}{2(\tilde{\gamma}-1)(1-\gamma)L}\Delta.\label{eq:eta}
\end{equation}
When $\Delta\rightarrow0$, $\mathcal{F}_{k^{*}}\rightarrow0$ and
thus $\ln\mathcal{F}_{k^{*}}$ is divergent, i.e., when $\phi$ achieves
$\phi_{c}$, $\alpha$ becomes divergent.
\begin{figure}
\begin{centering}
\includegraphics[scale=0.58]{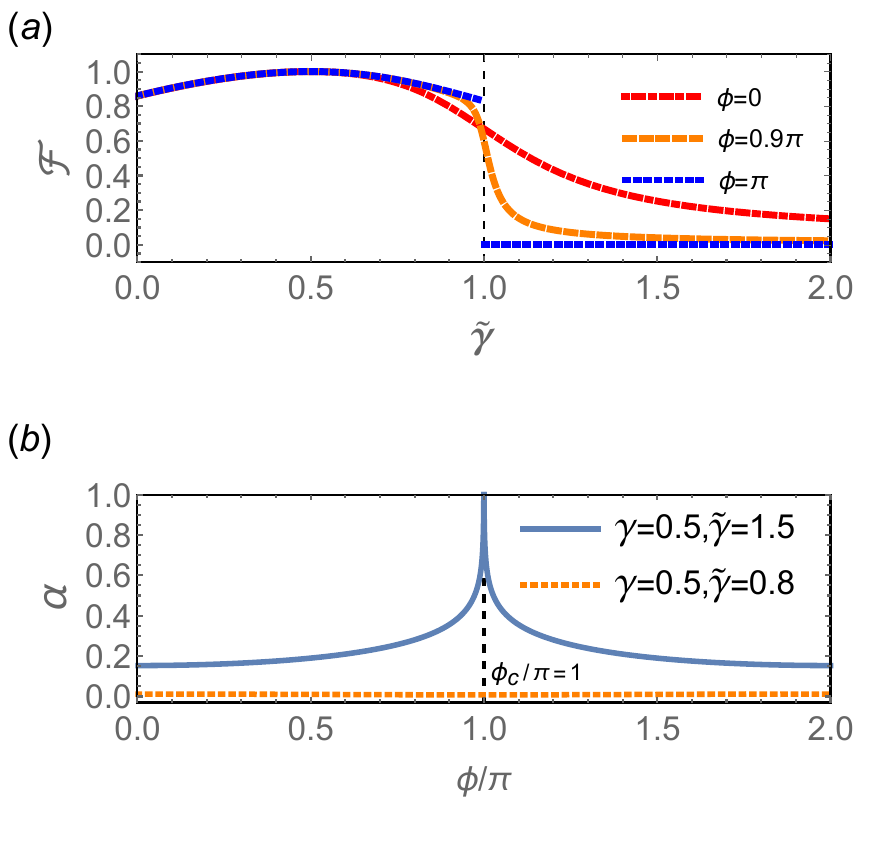}
\par\end{centering}
\caption{(a) The fidelity $\mathcal{F}$ of the SSH model versus $\tilde{\gamma}$
under twist boundary conditions with flux parameter $\phi=0$, $0.9\pi$,
and $\pi$. The vertical dashed line guides the critical point $\tilde{\gamma}_{c}=1$.
Here $\gamma=0.5$. (b) The images of $\alpha$ versus $\phi/\pi$.
The solid blue line corresponds to $\gamma=0.5$ and $\tilde{\gamma}=1.5$,
while the dotted orange line corresponds to $\gamma=0.5$ and $\tilde{\gamma}=0.8$.
The vertical dashed line guides the divergent point $\phi_{c}/\pi=1$.
Here we take $L=9$. \label{Fig3}}
\end{figure}

In Fig. \ref{Fig3}(a), we plot the fidelity $\mathcal{F}$ of the
SSH model versus $\tilde{\gamma}$ for different boundary conditions
$\phi=0$, $0.9\pi$ and $\pi$ with $\gamma=0.5$ and $L=9$. It
is evident that $\mathcal{F}$ immediately equals zero when $\tilde{\gamma}$
is across the critical point $\gamma_{c}=1$ for $\phi=\pi$, while
$\mathcal{F}$ stays non-zero for other $\phi$. The existence of the
exact zero of $\mathcal{F}$ for a finite odd size system means that
the ground state of one phase can be orthogonal to the ground state
of the other phase by tuning the magnetic flux. At the same time,
$\mathcal{F}_{k_{c}}=0$ leads to the divergency of $\alpha$ for
a finite $L$. For a given $\tilde{\gamma}$ and $\gamma$, tuning
$\phi$ from 0 to $2\pi$, from Fig. \ref{Fig3}(b) we can see
that if $\tilde{\gamma}$ and $\gamma$ belong to the same phase,
the value of $\alpha$ is always small for any value of $\phi$, indicating
the absence of the exact zero of the fidelity; if $\tilde{\gamma}$ and $\gamma$
belong to different phases, $\alpha$ is divergent at $\phi_{c}/\pi=1$,
which gives a signal of quantum phase transition. As a consequence,
we can judge whether a quantum phase transition exists by observing
the change of $\alpha$ of a finite system as a function of $\tilde{\gamma}$
and $\phi$.

\subsection{Creutz model}

Next we consider the Creutz model \cite{Creutz1999,Jafari} described
by the Hamiltonian:
\begin{align}
H & =-\sum_{j=1}^{L}[J_{h}(e^{i\theta}c_{j+1}^{p\dagger}c_{j}^{p}+e^{-i\theta}c_{j+1}^{q\dagger}c_{j}^{q})\nonumber \\
 & \qquad\qquad+J_{d}(c_{j+1}^{p\dagger}c_{j}^{q}+c_{j+1}^{q\dagger}c_{j}^{p})+J_{v}c_{j}^{q\dagger}c_{j}^{p}+\mathrm{H.c.}].
\end{align}
The model describes the dynamics of a spinless electron moving in
a ladder system with $c_{j}^{p(q)\dagger}$ and $c_{j}^{p(q)}$ denoting
fermionic creation and annihilation operators on the $j$-th site
of the lower (upper) chain. $J_{h}$, $J_{d}$ and $J_{v}$ are coupling
strength for horizontal, diagonal and vertical bonds, respectively;
and $\theta\in[-\pi/2,\pi/2]$ denotes the magnetic flux per plaquette
induced by a magnetic field piercing the ladder. After a Fourier transformation, the Hamiltonian in the momentum space can be expressed as $H= \sum_{k} \psi_k^{\dagger} \hat{h}_k \psi_k$
with $\psi_k =(c_{k}^{q},c_{k}^{p})^{T}$.
The vector components of the Hamiltonian in momentum space are given
by
\begin{eqnarray}
d_{x,k} & = & -2J_{d}\cos k-J_{v},\\
d_{z,k} & = & -2J_{h}\sin k\sin\theta,\\
d_{0,k} & = & -2J_{h}\cos k\cos\theta,
\end{eqnarray}
and $d_{y,k}=0$. For simplicity, we focus on the case of $J_{h}=J_{d}=J$
and $J_{v}/2J<1$, and take $J=1$ as the energy unit. It is known
that the Creutz model has two distinct topologically nontrivial phases
for $-\pi/2\leq\theta<0$ and $0<\theta\leq\pi/2$ with a phase transition
occurring at $\theta_{c}=0$.
\begin{figure}
\begin{centering}
\includegraphics[scale=0.47]{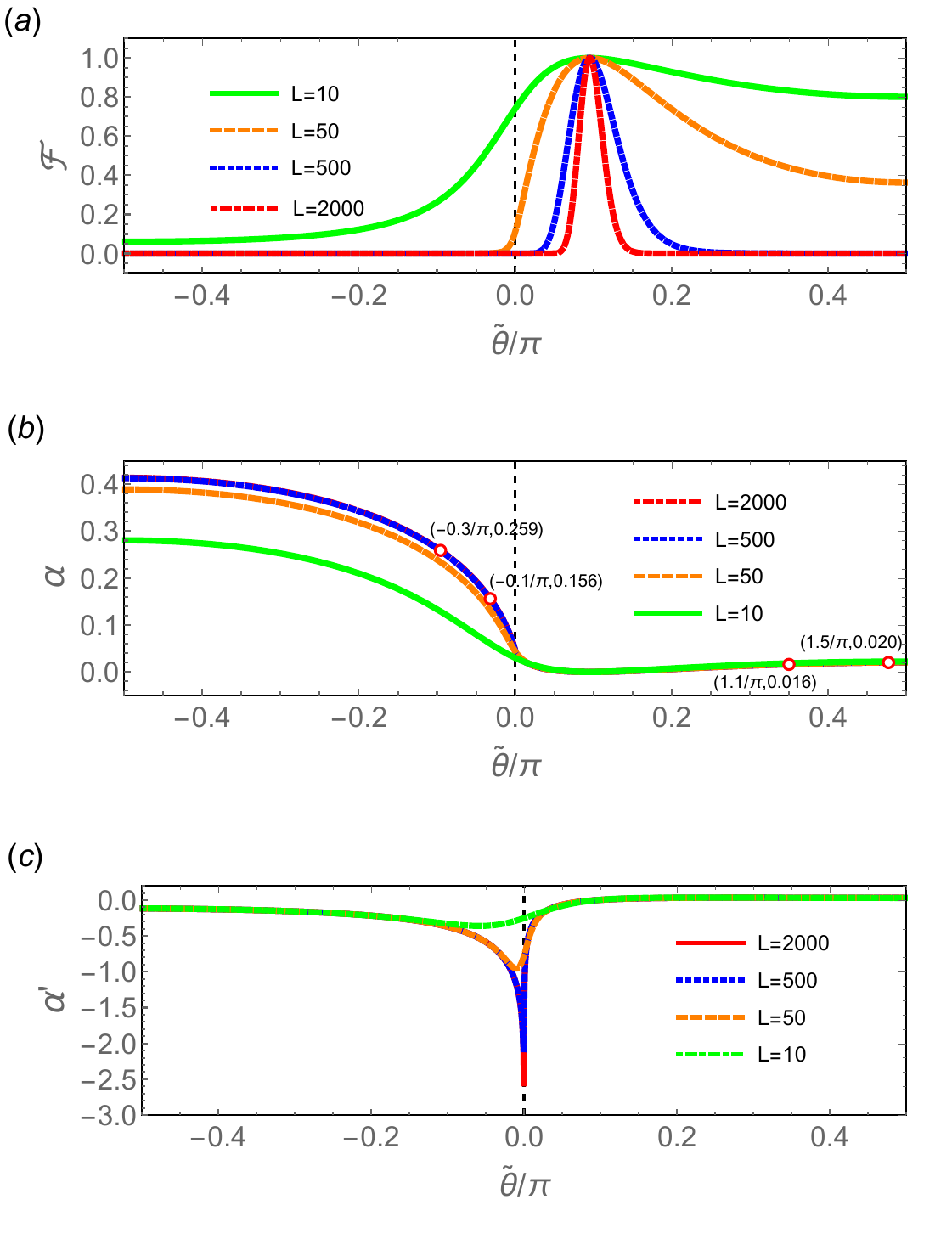}
\par\end{centering}
\caption{(a) The fidelity $\mathcal{F}$ of the Creutz model versus $\tilde{\theta}/\pi$
for different system sizes $L=10$, $50$, $500,$ and $2000$. The
vertical dashed line guides the critical point $\tilde{\theta}_{c}=0$.
(b) The images of $\alpha$ versus $\tilde{\theta}/\pi$
for the Creutz model with different system sizes $L=10$, $50$, $500$,
and $2000$. The four red circles denote valus of $\alpha$ in the
thermodynamical limit obtained from finite-size analysis for $\tilde{\theta}=-0.3,\ -0.1,\ 1.1,$
and $1.5$, respectively. (c) The images of the derivative of $\alpha$
versus $\tilde{\theta}/\pi$ for the Creutz model with
different system sizes $L=10$, $50$, $500$, and $2000$. The vertical
dashed line guides the critical point $\tilde{\theta}_{c}=0$. Here
we take $\theta=0.3$ and $J_{v}/(2J)=0.6$, and use the periodic boundary condition. \label{Fig4}}
\end{figure}

The fidelity of the Creutz model is given by
\begin{equation}
\mathcal{F}=\prod_{k}\sqrt{\frac{[\cos k+J_{v}/(2J)]^{2}+\sin^{2}k\sin\theta\sin\tilde{\theta}}{2\epsilon_{k}\tilde{\epsilon}_{k}}+\frac{1}{2}},
\end{equation}
where $\epsilon_{k}=\sqrt{[\cos k+J_{v}/(2J)]^{2}+\sin^{2}k\sin^{2}\theta}$
and $\tilde{\epsilon}_{k}=\sqrt{[\cos k+J_{v}/(2J)]^{2}+\sin^{2}k\sin^{2}\tilde{\theta}}$.
We notice that the constraint relations for ensuring $\mathcal{F}=0$
are
\begin{equation}
\sin\theta\sin\tilde{\theta}<0,\ k_{c,\pm}=\pm\arccos\left[-J_{v}/(2J)\right].\label{eq:thetaf}
\end{equation}
If $\theta<0$, Eq. (\ref{eq:thetaf}) is fulfilled only for $\tilde{\theta}>0$.
On the other hand, if $\theta>0$, Eq. (\ref{eq:thetaf}) is fulfilled
only for $\tilde{\theta}<0$. It means that the exact zeros of the
fidelity exist only when parameters $\tilde{\theta}$ and $\theta$
are across the underlying phase transition point and the corresponding
$k$ mode is $k_{c,\pm}=\pm\arccos\left[-J_{v}/(2J)\right]$.

It is clear that $k_{c,\pm}$ are usually not equal to the quantized
momentum values $k=2\pi m/L$ enforced by the PBC. This means that
the exact zeros of $\mathcal{F}$ of a finite-size system generally
do not exist for arbitrary $\tilde{\theta}$ and $\theta$. As shown
in Fig. \ref{Fig4}(a), the fidelity $\mathcal{F}$ is not equal to
zero for $L=10$, but approaches zero with the increase of $L$ as
long as $\tilde{\theta}\neq\theta$. Figure \ref{Fig4}(b) displays
the images of $\alpha$ versus $\tilde{\theta}$ for the Creutz model
with different system sizes $L=10$, $50$, $500$ and $2000$. For
$\tilde{\theta}$ and $\theta$ in the same phase, $\alpha$ is almost
the same for different sizes. For $\tilde{\theta}$ and $\theta$
in different phases, since it has two $k_{c}$ modes $\mathcal{F}_{k_{c,+}}$and
$\mathcal{F}_{k_{c,-}}$, the value of $\alpha$ changes and tends
to a fixed value at a rate of $2\ln2/L$ as the system size increases.
Figure \ref{Fig4}(c) demonstrates that $\alpha'$ is divergent at
the phase transition point $\theta_{c}=0$ in the thermodynamical
limit of $L\rightarrow\infty$.

For a finite-size system under PBC, $k_{c,\pm}$ is usually not achievable.
Nonetheless, with the increase in the system size, $k_{c,\pm}$ can
be approached in terms of $\min|k-k_{c,\pm}|\leq\pi/L$, and thus
the exact zeros of $\mathcal{F}$ can be achieved in the limit of
$L\rightarrow\infty$. Since the quantized momenta $k$ usually do
not include $k_{c,\pm}$ under the PBC, we introduce the twist boundary
condition $c_{L+1}^{p(q)\dagger}=c_{1}^{p(q)\dagger}e^{i\phi}(\phi\in(0,2\pi))$
here. For a system with a given finite size $L$, we can always achieve
$k_{c,+}$ or $k_{c,-}$ by using the twist boundary condition with
\begin{equation}
\phi_{c,+}=\!\negthinspace\!\!\!\!\mod\!\!\left[Lk_{c,+},2\pi\right]\ \mathrm{or}\ \phi_{c,-}=\!\negthinspace\!\!\!\!\mod\!\!\left[Lk_{c,-},2\pi\right].\label{Creutz-phic}
\end{equation}
For $J_{v}/(2J)=0.6$ and $L=10$, it is easy to get $k_{c,\pm}=\pm0.705\pi$,
$\phi_{c,+}\approx1.048\pi$ and $\phi_{c,-}\approx0.952\pi$\textcolor{red}{{}
}from Eqs. (\ref{eq:thetaf}) and (\ref{Creutz-phic}).

Let $\Delta=\left|\phi-\phi_{c,\pm}\right|$, we can get
\begin{equation}
\alpha=-\frac{1}{L}(\ln\mathcal{F}_{k^{*}}+\sum_{k\neq k^{*}}\ln\mathcal{F}_{k}),
\end{equation}
where $\mathcal{F}_{k^{*}}$ comes from the contribution of the $k^{*}$-mode
which is closest to $k_{c,\pm}$, i.e. $\left|k^{*}-k_{c,\pm}\right|=\Delta/L$.
Let $\Delta\rightarrow0$, we can get
\begin{equation}
\mathcal{F}_{k^{*}}\approx-\frac{\left|\sin\theta-\sin\tilde{\theta}\right|}{2\sin\theta\sin\tilde{\theta}L}\Delta,\label{eq:eta-1}
\end{equation}
It means when $\Delta\rightarrow0,$ i.e. $\phi\rightarrow\phi_{c}$,
$\mathcal{F}_{k^{*}}\propto \Delta$. When $\phi$ reaches
$\phi_{c,\pm}$, we can get a $k^{*}$-mode which satisfies $k^{*}=k_{c,\pm}$
and $\mathcal{F}_{k^{*}}=0$, thus $\alpha$ becomes divergent.
\begin{figure}
\begin{centering}
\includegraphics[scale=0.51]{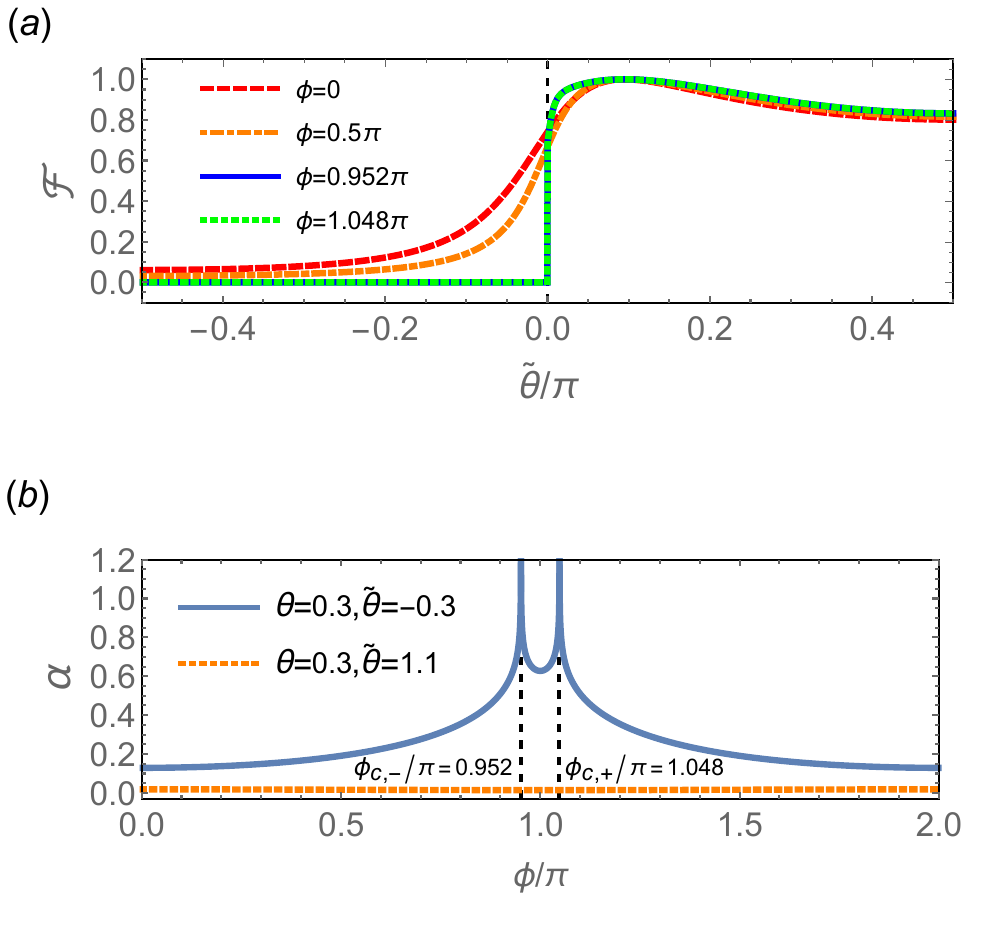}
\par\end{centering}
\caption{(a) The fidelity $\mathcal{F}$ of the Creutz model versus $\tilde{\theta}/\pi$
under twist boundary conditions with $\phi=0$, $0.5\pi$, $0.952\pi$
and $1.048\pi$. The solid blue line and the dotted green line
have the same image and therefore overlap each other. The vertical
dashed line guides the critical point $\tilde{\theta}_{c}=0$. Here
$\theta=0.3.$ (b) The images of $\alpha$ versus $\phi/\pi$. The
solid blue line corresponds to $\theta=0.3$ and $\tilde{\theta}=-0.3$,
while the dotted orange line corresponds to $\theta=0.3$ and $\tilde{\theta}=1.1$.
The vertical dashed lines guides the divergent points $\phi_{c,-}/\pi\approx0.952$ and $\phi_{c,+}/\pi\approx1.048$.
Here we take $J_{v}/(2J)=0.6$ and $L=10$. \label{Fig5}}
\end{figure}

In Fig. \ref{Fig5}(a), we show the fidelity $\mathcal{F}$ of the
Creutz model versus $\tilde{\theta}$ for different boundary conditions
$\phi=0$, $0.5\pi$, $0.952\pi$ and $1.048\pi$ with $\theta=0.3$,
$J_{v}/(2J)=0.6$, and $L=10$. It demonstrates that $\mathcal{F}$
becoms $0$ for $\tilde{\theta}<0$ when $\phi$ is tuned to the critical
value\textcolor{red}{{} }$\phi_{c,-}\approx0.952\pi$ and $\phi_{c,+}\approx1.048\pi$.
For a pair of given $\tilde{\theta}$ and $\theta$, Fig. \ref{Fig5}(b)
exhibits that if $\tilde{\theta}$ and $\theta$ belong to the same
phase, $\alpha$ barely changes with $\phi$, which means the absence
of QPT; if $\tilde{\theta}$ and $\theta$ belong to different phases,
$\alpha$ diverges at\textcolor{red}{{} }$\phi_{c,-}\approx0.952\pi$
and $\phi_{c,+}\approx1.048\pi$, indicating the occurrence of QPT.

\subsection{Haldane model}

\begin{figure}[H]
\begin{centering}
\includegraphics[scale=0.6]{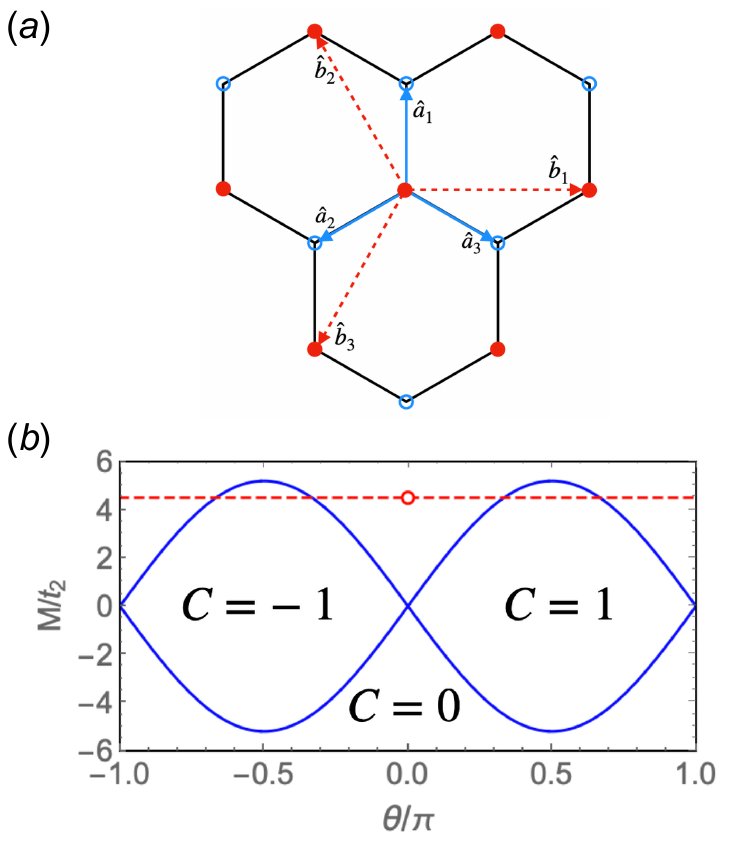}
\par\end{centering}
\caption{(a) The illustration of the Haldane model. (b) The phase diagram of the Haldane
model. The solid blue lines represent the critical lines. The dashed red line represents parameters at $M/t_{2}=4.5$. The red circle
denotes the point $(\theta/\pi,M/t_{2})=(0,4.5)$. Here we take $t_{1}=1$.
\label{Fig6}}
\end{figure}

The Haldane model is schematically depicted in Fig. \ref{Fig6}(a).
The \textcolor{red}{red points and blue circles} denote $A$ and $B$ sublattice sites, respectively.
The displacements are $\hat{a}_{1}=(0,1),\ \hat{a}_{2}=(-\frac{\sqrt{3}}{2},-\frac{1}{2})$,
$\hat{a}_{3}=(\frac{\sqrt{3}}{2},-\frac{1}{2})$, $\hat{b}_{1}=(\sqrt{3},0)$,
$\hat{b}_{2}=(-\frac{\sqrt{3}}{2},\frac{3}{2})$, and $\hat{b}_{3}=(-\frac{\sqrt{3}}{2},-\frac{3}{2})$.
The Hamiltonian of the Haldane model in the real space
is described as
\begin{align}
H= & t_{1}\sum_{\left\langle i,j\right\rangle }(c_{A,\vec{r}_{i}}^{\dagger}c_{B,\vec{r}_{j}}+\mathrm{H.c.})\nonumber \\
 & +t_{2}\sum_{\left\langle \left\langle i,j\right\rangle \right\rangle }(e^{-i\theta}c_{A,\vec{r}_{i}}^{\dagger}c_{A,\vec{r}_{j}}+e^{i\theta}c_{B,\vec{r}_{i}}^{\dagger}c_{B,\vec{r}_{j}}+\mathrm{H.c.})\nonumber \\
 & +M\sum_{j}(c_{A,\vec{r}_{j}}^{\dagger}c_{A,\vec{r}_{j}}-c_{B,\vec{r}_{j}}^{\dagger}c_{B,\vec{r}_{j}}).
\end{align}
Here $c_{A(B),\vec{r}_{j}}^{\dagger}$ and $c_{A(B),\vec{r}_{j}}$
denote fermionic creation and annihilation operators of $A(B)$ sublattice
on the position $\vec{r}_{j}$, $M$ $(-M)$ is the on-site potential
on $A(B)$ sublattice sites, symbols $\left\langle i,j\right\rangle $
and $\left\langle \left\langle i,j\right\rangle \right\rangle $ denote
nearest-neighbor (NN) and next-nearest-neighbor (NNN) hoppings, $\theta$
is the additional effective phase of hopping between NNN sites, $t_{1}$
and $t_{2}$ are amplitudes of NN hopping and NNN hopping, respectively.
Taking the periodic boundary condition along the $x$-axis and $y$-axis
and using the Fourier transformation, the Hamiltonian of the Haldane
model in the momentum space can be expressed as $H= \sum_{\mathbf{k}} \psi_\mathbf{k}^{\dagger} \hat{h}_\mathbf{k} \psi_\mathbf{k}$,
where $\psi_\mathbf{k} =(c_{\mathbf{k},A},c_{\mathbf{k},B})^{T}$ and $\hat{h}_{\mathbf{k}}$ is
given by \cite{Haldane}
\begin{align}
\hat{h}_{\mathbf{k}}= & 2t_{2}\cos\theta\sum_{i}\cos(\mathbf{k}\cdot\hat{b}_{i})\mathrm{I}\nonumber \\
 & +t_{1}\sum_{i}\cos(\mathbf{k}\cdot\hat{a}_{i})\sigma_{x}+t_{1}\sum_{i}\sin(\mathbf{k}\cdot\hat{a}_{i})\sigma_{y}\nonumber \\
 & +[M-2t_{2}\sin\theta\sum_{i}\sin(\mathbf{k}\cdot\hat{b}_{i})]\sigma_{z},
\end{align}
where $i=1,2,3,$ $\mathbf{k}=(k_{x},k_{y})$ with $k_{x}=\frac{2\pi m_{x}}{\sqrt{3}L_{x}}$
$(m_{x}=1,2,\cdots,L_{x})$ and $k_{y}=\frac{4\pi m_{y}}{3L_{y}}$
$(m_{y}=1,2,\cdots,L_{y})$. It is easy to get
\begin{eqnarray*}
d_{x,\mathbf{k}}  &=&t_{1}(\cos k_{y}+2\cos\frac{\sqrt{3}k_{x}}{2}\cos\frac{k_{y}}{2}), \\
d_{y,\mathbf{k}}  &=& t_{1}(\sin k_{y}-2\cos\frac{\sqrt{3}k_{x}}{2}\sin\frac{k_{y}}{2}), \\
d_{z,\mathbf{k}}  &=& M-2t_{2}\sin\theta(\sin\sqrt{3}k_{x}-2\sin\frac{\sqrt{3}k_{x}}{2}\cos\frac{3k_{y}}{2}),
\end{eqnarray*}
and
\[
d_{0,\mathbf{k}} =2t_{2}\cos\theta(\cos\sqrt{3}k_{x}+2\cos\frac{\sqrt{3}}{2}k_{x}\cos\frac {3}{2}k_{y}).
\]

Fig. \ref{Fig6}(b) shows the phase diagram of the Haldane model.
The critical lines of the Haldane model are described by $M/t_{2}=\pm3\sqrt{3}\sin\theta$,
which separate topologically different phases characterized by the
Chern number with $C=0$ or $C=\pm1$ \cite{Haldane}. The corresponding
constraint relations for the occurrence of zeros of fidelity are
\begin{widetext}
\begin{equation}
\begin{cases}
(M/t_{2}+3\sqrt{3}\sin\theta)(M/t_{2}+3\sqrt{3}\sin\tilde{\theta})<0, & \begin{array}{c}
k_{xc}=\frac{4\pi}{3\sqrt{3}},k_{yc}=\frac{4\pi}{3};\end{array}\\
(M/t_{2}-3\sqrt{3}\sin\theta)(M/t_{2}-3\sqrt{3}\sin\tilde{\theta})<0, & \begin{array}{c}
k_{xc}=\frac{2\pi}{3\sqrt{3}},k_{yc}=\frac{2\pi}{3}.\end{array}
\end{cases}\label{eq:Haldane}
\end{equation}
\end{widetext}

According to the above constraint relation, if we choose $\theta=0,\ t_{1}=1,\ M/t_{2}=4.5,$
zeros of fidelity can be accessible in the thermodynamic limit for
$\tilde{\theta}\in(-2\pi/3,-\pi/3)\cup(\pi/3,2\pi/3)$, which is in
accordance with the red dashed line in Fig. \ref{Fig6}(b). Next we
consider the finite-size system. For $\tilde{\theta}\in(-2\pi/3,-\pi/3)$,
$k_{yc}=\frac{4\pi}{3}$ is always accessible. For $\tilde{\theta}\in(\pi/3,2\pi/3)$,
$k_{yc}=\frac{2\pi}{3}$ is accessible for even $L_{y}$, while a
twist boundary condition $c_{A(B),\vec{r}_{j}+L_{y}(\hat{b}_{2}-\hat{b}_{3})/2}^{\dagger}=c_{A(B),\vec{r}_{j}}^{\dagger}e^{i\phi_{y}}$
along the $y$-direction is needed such that $k_{y}=\frac{2(2\pi m_{y}+\phi_{yc})}{3L_{y}}$
with $\phi_{yc}=\pi$ can give rise to $k_{yc}=\frac{2\pi}{3}$ for
odd $L_{y}$. For simplicity, we choose $L_{y}=2n$ (even), so that
$k_{yc}$ is accessible with no need of the introduction of the twist
boundary condition along the $y$-direction.

\begin{figure}[H]
\begin{centering}
\includegraphics[scale=0.39]{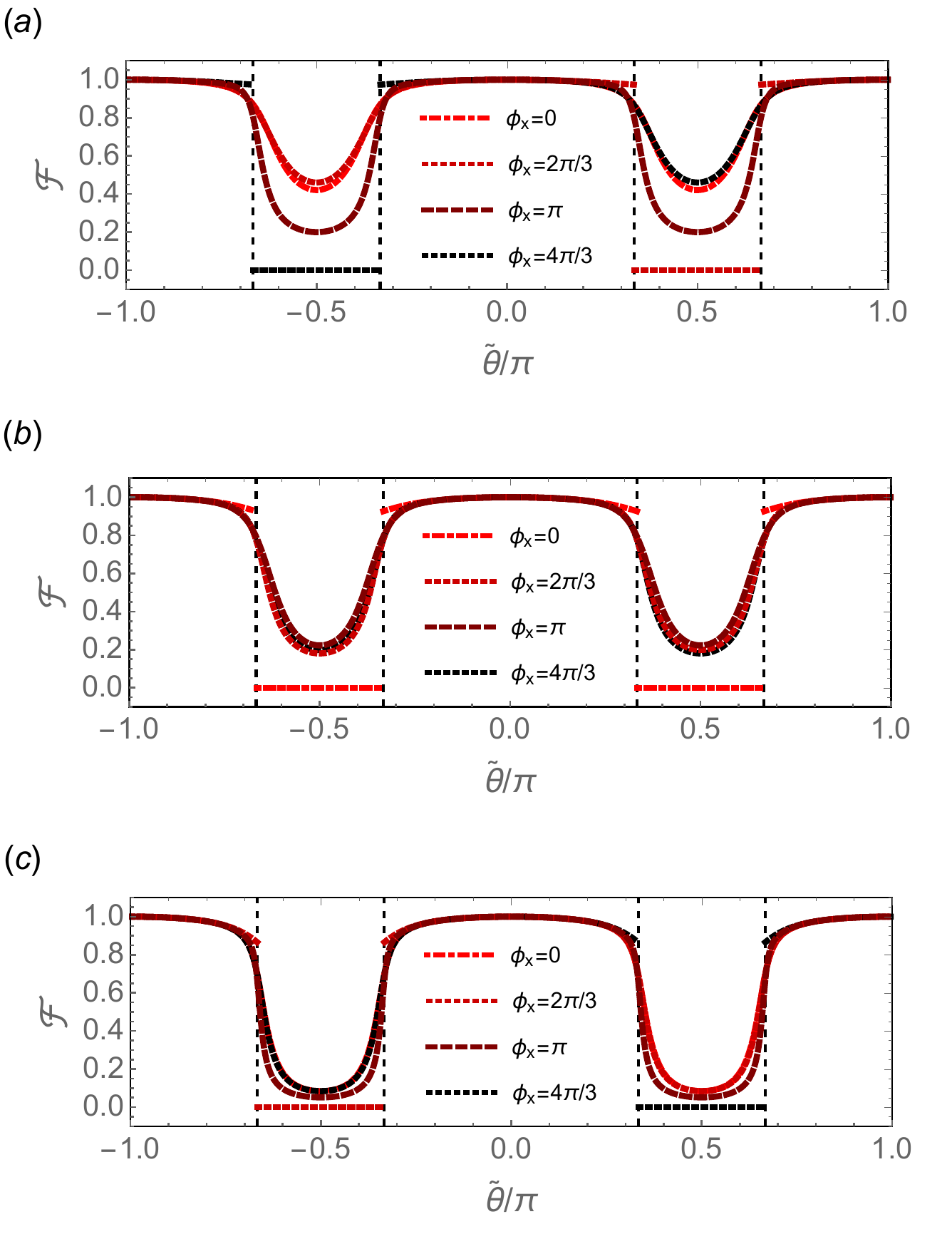}
\par\end{centering}
\caption{The fidelity $\mathcal{F}$ of the Haldane model versus $\tilde{\theta}/\pi$
under twist boundary conditions with various $\phi$. The vertical
dashed lines guides the critical points $\tilde{\theta}_{c}/\pi=\pm1/3,\pm2/3$.
(a) $L_{x}=4,\ L_{y}=4$. (b) $L_{x}=6,\ L_{y}=4$. (c) $L_{x}=8,\ L_{y}=4$.
Here we take $t_{1}=1,\ M/t_{2}=4.5,$ $\theta=0$.
\label{Fig7}}
\end{figure}

By applying the twist boundary condition $c_{A(B),\vec{r}_{j}+L_{x}\hat{b}_{1}}^{\dagger}=c_{A(B),\vec{r}_{j}}^{\dagger}e^{i\phi_{x}}$
along the $x$-direction, we demonstrate that exact zeros of $\mathcal{F}$
can be accessible by tuning the twist flux $\phi_{x}$ when $\tilde{\theta}\in(-2\pi/3,-\pi/3)\cup(\pi/3,2\pi/3)$.
In Fig. \ref{Fig7}(a), we show the images of the fidelity $\mathcal{F}$
of the Haldane model versus $\tilde{\theta}$ for systems with different
twist flux, where $\theta=0,\ t_{1}=1,\ M/t_{2}=4.5,$ $L_{x}=4$,
and $L_{y}=4$. It is shown that $\mathcal{F}$ drops to zero abruptly
at the points $\tilde{\theta}=-2\pi/3$ and $\tilde{\theta}=-\pi/3$
for $\phi_{x}=4\pi/3$ and at the points $\tilde{\theta}=\pi/3$ and
$\tilde{\theta}=2\pi/3$ for $\phi_{x}=2\pi/3$, while the fidelity
for other $\phi_{x}$ is analytic everywhere. Under the twist boundary
condition, the momentum $k_{x}$ is shifted and we have $\mathbf{k}=(k_{x}=\frac{2\pi m_{x}+\phi_{x}}{\sqrt{3}L_{x}},k_{y}=\frac{4\pi m_{y}}{3L_{y}})$.
For $\tilde{\theta}\in(\pi/3,2\pi/3)$, the $\mathbf{k}_{c}$ mode
can be accessed by tuning the twist flux to $\phi_{xc,1}=2\pi/3$,
whereas for $\tilde{\theta}\in(-2\pi/3,-\pi/3)$, the $\mathbf{k}_{c}$
mode can be accessed by tuning the twist flux to $\phi_{xc,2}=4\pi/3$.
Actually, for all cases of $L_{x}=3n+1$ with $n$ being a positive
integer, we have $\phi_{xc,1}=2\pi/3$ for $\tilde{\theta}\in(\pi/3,2\pi/3)$
and $\phi_{xc,2}=4\pi/3$ for $\tilde{\theta}\in(-2\pi/3,-\pi/3)$.
For cases with $L_{x}=3n$, we have $\phi_{xc}=0$. For cases with
$L_{x}=3n+2$, we have $\phi_{xc,1}=4\pi/3$ for $\tilde{\theta}\in(\pi/3,2\pi/3)$
and $\phi_{xc,2}=2\pi/3$ for $\tilde{\theta}\in(-2\pi/3,-\pi/3)$.
To see it clearly, we show the case of $L_{x}=6$ and $L_{x}=8$ in
Figs. \ref{Fig7}(b) and \ref{Fig7}(c), respectively. Although the
value of $\phi_{xc}$ may depend on the size of the system, it is
clear that the exact zero of the fidelity can be always accessed by continuously
tuning the twist flux $\phi_{x}$ as long as $\tilde{\theta}$ and
$\theta$ are in different phase regions.

\subsection{Interacting SSH model}
To exhibit the applicability of our theoretical scheme, here we study
one more example by taking into account interaction. We shall explore
the fidelity of the SSH model with interaction. The Hamiltonian of
the interacting SSH model is
\begin{align}
H & =\sum_{j=1}^{L}(c_{j,A}^{\dagger}c_{j,B}+\gamma c_{j,B}^{\dagger}c_{j+1,A}+\mathrm{H.c.})\nonumber \\
 & \ \ \ +U\sum_{j=1}^{L}(n_{j,A}n_{j,B}+n_{j,B}n_{j+1,A}),
\end{align}
where $U$ is the the magnitude of the interaction strength between
fermions on nearest-neighboring sites and $n_{j,A(B)}=c_{j,A(B)}^{\dagger}c_{j,A(B)}$.
Here we only consider the half-filling case. Ref.\cite{Tang} has
analyzed topological phase transitions in the interacting SSH model.
For large $U$, the system is in a density-wave phase, whereas for
small $U$, there is still a phase transition through varying the
parameter $\gamma$. We have obtained the approximate value of the
phase transition point $\gamma_{c}\approx1.038$ for $U=0.1$ through
the finite-size-scaling analysis of the fidelity. The numerical result
suggests that the transition point is close to $\gamma=1$.

\begin{figure}
\begin{centering}
\includegraphics[scale=0.6]{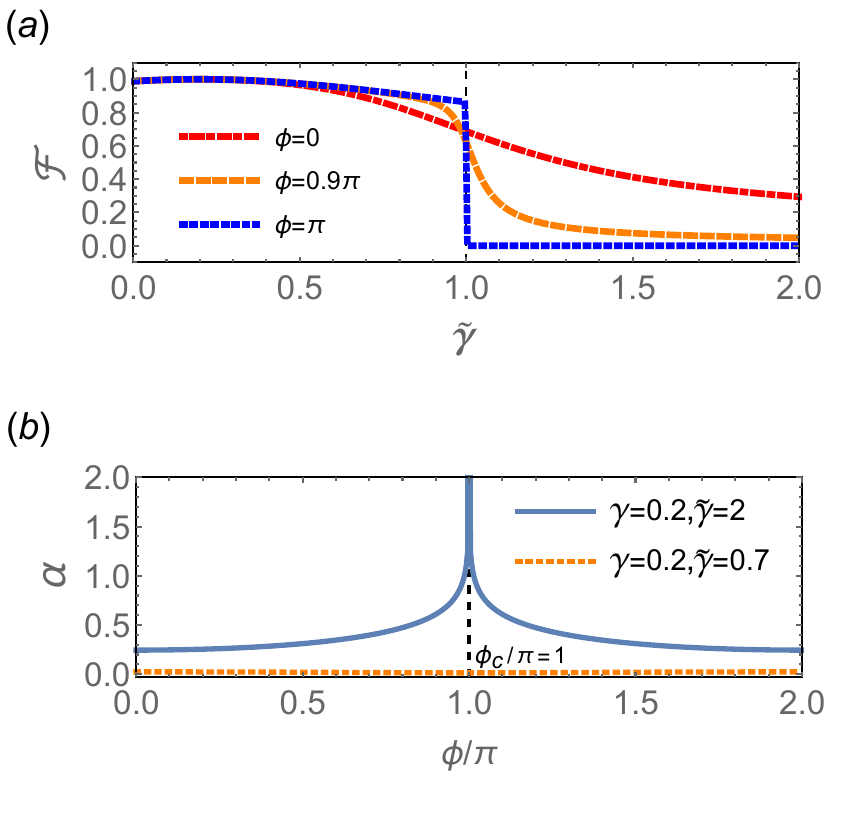}
\par\end{centering}
\caption{(a) The fidelity $\mathcal{F}$ of the interacting SSH model versus $\tilde{\gamma}$
under twist boundary conditions with $\phi=0$, $0.9\pi$, and $\pi$.
The vertical dashed line guides the critical point $\tilde{\gamma}_{c}=1$.
Here $\gamma=0.2$. (b) The images of $\alpha$ versus $\phi/\pi$.
The solid blue line corresponds to $\gamma=0.2$ and $\tilde{\gamma}=2$,
while the dotted orange line corresponds to $\gamma=0.2$ and $\tilde{\gamma}=0.7$.
The vertical dashed line guides the divergent point $\phi_{c}/\pi=1$.
Here we take $L=5$, $U=0.1$.\label{Fig8}}
\end{figure}

We numerically calculate the fidelity $F(\gamma,\tilde{\gamma})$
versus $\tilde{\gamma}$ via exact diagonalization of a system with
$L=5$ by fixing $\gamma=0.2$ for $U=0.1$ under the twist boundary
condition $c_{L+1,A(B)}^{\dagger}=c_{1,A(B)}^{\dagger}e^{i\phi}$
with various flux $\phi$. Our numerical results are displayed in
Fig. \ref{Fig8}. In Fig. \ref{Fig8}(a), for $\gamma=0.2$, we find
that the fidelity abruptly drops to zero at $\tilde{\gamma}=1$ only
under the antiperiodic boundary condition ($\phi=\pi$), suggesting
that the phase transition point is given by $\gamma_{c}=1$ for $U=0.1$.
In contrast, no sharp drop occurs and no exact zero of fidelity can
be obtained for other $\phi$. In Fig. \ref{Fig8}(b), the images
of $\alpha$ demonstrate that $\phi_{c}=\pi$ when $\tilde{\gamma}$
and $\gamma$ belong to different phases. Our numerical results indicate
that exact zeros of the fidelity obtained via the tuning of $\phi$ can
provide a clear signature of the quantum phase transition even by studying
a small size system of interacting SSH model.

\section{Summary and discussion}

In summary, we proposed a theoretical scheme for detecting quantum
phase transition by seeking exact zeros of fidelity
of finite-size systems with twist boundary conditions. By considering
the SSH model, the Creutz model and Haldane model as concrete examples,
we demonstrated that exact zeros of fidelity of finite-size
systems can be always accessed for $\tilde{\gamma}$ and $\gamma$
in different phases under a proper twist boundary condition, whereas
no exact zero exists for $\tilde{\gamma}$ and $\gamma$ in the same
phase. Consequently, we can observe a discontinued behavior of fidelity
at the phase transition point by tuning the twist flux parameter $\phi$.
Changing $\phi$ continuously, we unveiled that the decay rate function
$\alpha$ of the fidelity is divergent at the critical
magnetic flux $\phi_{c}$ for $\tilde{\gamma}$ and $\gamma$ in different
phases, while $\alpha$ is smooth everywhere for $\tilde{\gamma}$
and $\gamma$ in the same phase. We also exhibited the applicability
of our theoretical scheme to the interacting SSH model.

Our work provides
an efficient way for detecting quantum phase transition by studying
small-size systems via the introduction of an additional magnetic
flux. A natural question is whether such a scheme is applicable to a general context of models with quantum phase transitions, including correlated systems, the
Ising model with finite/infinite interaction range and models without a spatial
lattice interpretation, e.g., the Lipkin-Meshkov-Glick model \cite{LMG,LMG2}? Although we show the applicability of our theoretical scheme to the interacting SSH model in Sec.II D, our scheme does not always work for general correlated systems and models without spatial interpretation, for which the momentum is even not well defined. In the present work, the gap between the first excited state and ground state is a function of momentum for all the studied models, and thus we can tune the gap to approach zero by tuning the magnetic flux properly. For systems without spatial interpretation, the gap of a finite-size system is not necessary to be a function of momentum, and thus our scheme is not necessary to be applicable to these systems.

Our scheme also suggests an alternative way for experimentally
detecting signature of QPT in finite-size quantum systems, which however relies on the ability of creating tunable magnetic flux in quantum simulators.
For the experimental setup in a trapped-ion quantum simulator \cite{Monroe2017Nature,Jurcevic2017PRL},
it is still a great challenge  to create tunable magnetic flux in the
setup. However, it seems that cold atomic system is a promising platform for producing tunable magnetic flux. By using multifrequency Bragg lasers, the SSH
model on a momentum lattice was experimentally realized \cite{YanB}, where the synthetic magnetic
flux through the ring are tunable \cite{YanB2020}. Thus we expect it to be a promising platform
to observe signature of QPT by using the theoretical scheme proposed in this work.
\begin{acknowledgments}
The work is supported by National Key Research and Development Program
of China (Grant No. 2021YFA1402104), the NSFC under Grants No.12174436
and No.T2121001 and the Strategic Priority Research Program of Chinese
Academy of Sciences under Grant No. XDB33000000.

\end{acknowledgments}


\begin{thebibliography}{10}
\bibitem{note} The oblique line in the Fig. \ref{Fig1}(b) is fitted by  $\alpha=c - \ln2/L$, which
gives the lower bound of the $\alpha$ and can be derived by analytically
calculating the difference of the integral and the summation $-\frac{1}{2\pi}\int\ln\mathcal{F}_{k}\mathrm{d}k-(-\frac{1}{L}\sum_{k}\ln\mathcal{F}_{k})$.
For $\tilde{\gamma}$ and $\gamma$ in the same phase, $\ln \mathcal{F}_{k}$ is a smooth function of  $k$, and $-\frac{1}{2\pi}\int\ln\mathcal{F}_{k}\mathrm{d}k-(-\frac{1}{L}\sum_{k}\ln\mathcal{F}_{k})\ll 1/L $. For $\tilde{\gamma}$ and $\gamma$ in different phases, if the fidelity has a $k_{c}$-mode which is exactly situated in the middle
of two quantized momentum $k$-mode\textcolor{red}{s}, i.e. $\min|k-k_{c}|=\pi/L$,
then $-\frac{1}{2\pi}\int_{0}^{k_{c}-\pi/L} \ln\mathcal{F}_{k}\mathrm{d}k-\frac{1}{2\pi}\int_{k_{c}+\pi/L}^{2\pi} \ln\mathcal{F}_{k}\mathrm{d}k-(-\frac{1}{L}\sum_{k\ne k_{c}-\pi/L}\ln\mathcal{F}_{k})\ll 1/L$ and $-\frac{1}{2\pi}\int_{k_{c}-\pi/L}^{k_{c}+\pi/L} \ln\mathcal{F}_{k}\mathrm{d}k-(-\frac{1}{L}\ln\mathcal{F}_{k\textcolor{red}{_c}-\pi/L})\approx \ln2/L$. Adding them up, we get $-\frac{1}{2\pi}\int_{0}^{2\pi}\ln\mathcal{F}_{k}\mathrm{d}k-(-\frac{1}{L}\sum_{k}\ln\mathcal{F}_{k})\approx \ln2/L$. As a result, we get $\alpha(L)\approx\alpha_{\infty}-\ln2/L$, here $\alpha_{\infty}$
is the value of $\alpha$ in the thermodynamic limit. If the fidelity
has a $k_{c}$-mode satisfises $\min|k-k_{c}|<\pi/L$, then $\alpha(L)>\alpha_{\infty}-\ln2/L$.
The above is the case with the fidelity having only one $k_{c}$-mode.
For the case with the fidelity having $n$ $k_{c}$-modes, the slope
of the lower bound is $-n\ln2$.
\end{thebibliography}
\end{document}